\documentclass[11pt]{article}
 \usepackage{amsfonts}
  \usepackage{amssymb}

\makeatletter

 \def\nhline{\noalign{\ifnum0=`}\fi\hrule \@height \arrayrulewidth 
 \futurelet
   \@tempa\@xhline}
 \def\phantomthreeptline{\noalign{\ifnum0=`}\fi\vskip 3pt 
 \futurelet
   \@tempa\@xhline}

\makeatother

 \def\bsh{\backslash}
 \def\bdt{\dot \beta}
 \def\adt{\dot \alpha}

\newfont{\goth}{eufm10 scaled \magstep1}

 \def\gu{\mbox{\goth u}}

 \def\be{\begin{equation}}\def\ee{\end{equation}}
 \def\bea{\begin{eqnarray}}\def\eea{\end{eqnarray}}
 \def\ba{\begin{array}}\def\ea{\end{array}}
 \def\del{\partial}
 \def\xz{\times}
 
 \def\nn{\nonumber}
 \def\bea{\begin{eqnarray}}\def\barr{\begin{array}}\def\earr{\end{array}}
 \def\eea{\end{eqnarray}}
 \def\ft#1#2{{\textstyle{{\scriptstyle #1}\over {\scriptstyle #2}}}}

 \newcommand{\eq}[1]{(\ref{#1})}

 \def\bm{\bibitem}
 \def\nl{\newline}

\begin{document}

 \thispagestyle{empty}

\begin{flushright}
\hfill{KCL-MTH-02-27}\\
\hfill{Imperial/TP/1-02-03/6}\\
\hfill{hep-th/0211279}
\vskip .5cm
\hfill{November, 2002}
\end{flushright}

 \vspace{20pt}

 \begin{center}
 {\Large{\bf Supersymmetry counterterms revisited}}
 \vspace{30pt}

 {P.S. Howe} \vskip .2cm {Department of Mathematics}\linebreak
 {King's College, Strand, London WC2R 2LS}
  \vskip .3cm
  {and}
  \vskip .3cm
 {K.S. Stelle} \vskip .2cm{Theoretical Physics Group},
 \linebreak {Imperial College, Prince Consort Road, London SW7 2BW} \vspace {15pt}
  \vspace{40pt}
 \end{center}

\begin{abstract}
\normalsize
Superspace power-counting rules give estimates for the loop order
at which divergences can first appear in non-renormalisable
supersymmetric field theories. In some cases these estimates can be
improved if harmonic superspace, rather than ordinary superspace,
is used. The new estimates are in agreement with recent results
derived from unitarity calculations for maximally supersymmetric
Yang-Mills theories in five and six dimensions. For $N=8$
supergravity in four dimensions, we speculate that the onset of
divergences may correspondingly occur at the six loop order.
\end{abstract}

 \baselineskip=15pt \pagebreak \setcounter{page}{1}


\section{Introduction}

In this note we review the question of the onset of ultra-violet
divergences in super Yang-Mills theories (in more than four
dimensions) and in supergravity. One of the most distinctive features
of supersymmetric quantum field theories is that they generally
have improved ultra-violet behaviour compared to their
non-supersymmetric counterparts \cite{early}. This became evident shortly after
the discovery of four-dimensional supersymmetry and all-orders
results were subsequently established for renormalisable theories.
These results are most clearly seen in the framework of superspace
where non-renormalisation theorems can be established using
superspace power-counting \cite{griegelcek}. This approach can also be applied to
non-renormalisable theories, in particular to supergravity and
higher-dimensional super Yang-Mill theories (SYM). Here, although
early hopes that the maximally supersymmetric $N=8$ supergravity
might be finite do not seem to be justifiable
\footnote{3-loop counterterms were constructed for $N=1,2$ supergravities in \cite{dks}
and a 7-loop counterterm was found for $N=8$ supergravity in \cite{hl}.}, superspace
techniques can nevertheless be used to predict the lowest
loop-order at which the onset of ultra-violet divergences can be
expected to occur. Typically, the onset of divergences occurs at
higher loop order than in non-supersymmetric theories, but no
non-renormalisable field theory is rendered finite by
supersymmetrisation.

In \cite{miraculous,hs} predictions were made for the onset of divergences in
supergravity and higher-dimen\-sional SYM theories, using techniques
from \cite{grisetal,relaxed}. The status of these original predictions based upon
standard superspace was reviewed in \cite{hsrev}. These predictions were based on two
assumptions: firstly, that the counterterm corresponding to the lowest loop-order
divergence should be invariant under the full ``on-shell'' $N$-extended
supersymmetry\footnote{In this paper, we refer to minimal supersymmetry in a given
dimension as $N=1$ supersymmetry, {\it etc.} For clarity, we also indicate, when
appropriate, the corresponding number of supercharges.} of the theory and secondly, that
it should be expressible as a full superspace integral with respect to the portion $M$ of
the full supersymmetry that is linearly realisable ``off-shell'' and so can be preserved
manifestly in the quantisation procedure.

The non-renormalisation theorem of Refs \cite{grisetal,miraculous} is
conveniently expressed using the background field method,\footnote{It can also 
 be derived straightforwardly using standard superspace Feynman rules.} and draws its power
from the fact that, in a background-quantum split, only certain combinations of the
background fields appear in the Feynman rules used in calculating corrections to the
quantum effective action $\Gamma$ (the generating functional of 1PI diagrams). Although
it is convenient to solve superspace constraints to express the {\em quantum} fields
appearing on the internal lines of 1PI superspace Feynman diagrams as differential
constructions using unconstrained ``prepotentials,'' this is never necessary for
the {\em background} fields. The non-renormalisation theorem that follows from this is:
\begin{quote}
For gauge and supergravity multiplets at loops $\ell\geq 2$, and for matter
multiplets at all loop orders, counterterms must be written as full superspace
integrals for the maximal off-shell linearly realisable supersymmetry, with
background gauge invariant integrands written without using prepotentials for the
background fields.
\end{quote}
Additional restrictions on counterterms may also follow by combining the
non-renormalisation theorem with the full on-shell supersymmetry. The on-shell
supersymmetry is not as powerful as the off-shell, linearly-realisable supersymmetry
because it is non-linear and also has an algebra that closes only modulo
field-equation transformations, thus allowing poorly controllable mixing between
counterterms and the original action. However, imposing the original classical field
equations removes variations of the original action from consideration, and restriction
to the lowest order divergences leaves them nothing of lower order to mix with, so a
simple statement of invariance with respect to the full on-shell supersymmetry, modulo the
classical field equations, is obtained. This can have the effect of linking allowable
counterterms under the linearly realisable $M$ supersymmetry to counterterms that are
thereby disallowed, requiring thus an overall vanishing coefficient \cite{hs}.
Consequently, the simple requirement of invariance for the leading divergences under the
full on-shell supersymmetry modulo classical field equations is significantly
strengthened upon combination with the non-renormalisation theorem.

In practice, for the maximally supersymmetric Yang-Mills and
supergravity theories, with $16$ and $32$ supersymmetries
respectively, it was assumed\footnote{For $D=4$, $N=4$ SYM, a full
quantum formalism was given in Refs \cite{relaxed,miraculous} in terms of $D=4$, $M=2$
superfields (corresponding to 8 supercharges). Similarly, the $D=5, N=2$ and $D=6, N=2$ 
SYM theories can be written in terms of $M=1$ superfields in these dimensions 
\cite{hst,koller}.
Higher dimensional analogues of this
8-supercharge formalism would linearly realise only a partial Lorentz covariance; for
example, one could work with a formalism that linearly realises $D=6$ Lorentz covariance
and 8 supercharges in the maximal $D\geq 7$ SYM theory. For $D=4$, $N=8$ supergravity,
linearised analysis reveals the possibility of a $D=4$, $M=4$ ordinary superspace
formalism. Correspondingly, an off-shell version of linearised $D=10, M=1$ supergravity
was constructed in \cite{Howe:1982mt}. Multiplet counting arguments at the linearised
level \cite{rivelles&taylor} also suggest the existence of an $M=1$ off-shell version of
the
$D=10$ spin 3/2 multiplet that is needed to complete the maximal $D=10$, $N=2$ theory,
but the details of this have not been worked out.} in \cite{miraculous,hs} that one-half
of the supersymmetries can be preserved manifestly in the quantum theory; in other words,
that one can quantise these theories, at least in principle, using superfields with $8$ or
$16$ supersymmetries.

On the basis of these assumptions it was argued in \cite{miraculous,hsrev} that the lowest
counterterm would be of the form $R^4$ in supergravity (3 loops in $D=4$), while for $D=5$
SYM the first divergence should occur at 4 loops corresponding to
a counterterm of the form $F^4$. However, recent computations \cite{bernetal} using
unitarity cutting-rule techniques have indicated that the onset of divergences actually
occurs at higher loop order in the $D=5$ maximal SYM theory and also in $D=4$
maximal supergravity. Specifically, the authors of \cite{bernetal} find that
divergences start at 6 loops in $D=5$ SYM and that their onset is
postponed to at least 5 loops in the $D=4, N=8$ supergravity
theory.
 Previously it had been shown that the maximal SYM theories are finite at two loops 
in $D=5,6$ \cite{ms}, but the new unitarity calculations strengthen this result 
substantially for the $D=5$ case.

In \cite{karpacz} an attempt was made to explain these results using higher-dimensional
gauge-invar\-iance. If a theory in a given spacetime dimension somehow knows about the
gauge-invariances of the theory in higher dimensions from which it can be derived by
dimensional reduction, then one could expect to have improved ultra-violet divergence
behaviour because, for example, terms involving undifferentiated scalar fields would be
ruled out by the higher-dimensional gauge symmetry. In practice, this argument seems hard
to justify, and indeed does not seem to be true in simpler non-supersymmetric examples.
However, there is a simpler reason why one would expect to find better UV behaviour than
predicted in \cite{miraculous,hs,hsrev}, and that is that it may be possible to preserve
a larger fraction of the supersymmetry manifestly in the quantum theory that had been
assumed. The results of \cite{miraculous,hs,hsrev} depend on power-counting in ordinary
superspace, but one can obtain improved power-counting results if one makes us of
harmonic superspace \cite{Galperin:1984av}. 
Indeed, this formalism has been available since the early $1980$s
but has been  overlooked in the context of non-renormalisable theories.

It was shown in \cite{gikosN4} that the maximal SYM theory in $D=4$ admits an off-shell
superfield formulation (and therefore a quantisation procedure) in $M=3$ harmonic
superspace, so that one can therefore preserve 3/4 of the supersymmetries rather than
just 1/2. If one uses this formalism for higher-dimensional theories, even though the
manifest higher-dimensional Lorentz symmetry is lost, one finds that the expected onset
of divergences in $D=5$ now agrees with the results of \cite{bernetal}. The situation in
$D=6$, however, is unchanged, and indeed the old results for this theory are in agreement
with the new calculations. For the maximal supergravity theory it is not known how much
supersymmetry can be preserved off-shell using harmonic superspace methods, but the
similarity of the relationships between $D=4$ $N=3$ and $N=4$ SYM and $D=4$ $N=7$ and
$N=8$ supergravity lead us to conjecture that there may be an off-shell version of $N=8$
supergravity with $M=7$ supersymmetry. If this is true, then one would expect the onset
of divergences to occur at 6 loops, while if only $M=6$ supersymmetry can be preserved,
then the divergences would start at 5 loops.

\section{Yang-Mills theory}

The $D=4$, $N=4$ SYM theory is described in ordinary Minkowski superspace
by a scalar superfield $W_{ij}=-W_{ji},\,i,j=1\ldots 4$ which
transforms under the six-dimensional representation of the $SU(4)$
internal symmetry group, and which is also real in the sense that
$\bar W^{ij}=\ft12 \epsilon^{ijkl} W_{kl}$. It obeys the constraints
 \bea
 \nabla_{\alpha i} W_{jk}&=&\nabla_{\alpha [i} W_{jk]}\\
 \bar \nabla_{\adt}^i W_{jk}&=&-{2\over3}\delta_{[j}{}^i
 \bar \nabla_{\adt}^l W_{k]l}
 \label{dotW}
 \eea
where $(\nabla_{\alpha i},\bar\nabla_{\adt}^i)$ are gauge-covariant
spinorial derivatives. In fact, the reality constraint on $W_{ij}$
means that the above two constraints are equivalent. These equations
follow from the imposition of the standard constraints on the
superspace field-strength two-form; by use of the Bianchi
identities one can show that they imply that the only component fields are those
of the on-shell SYM multiplet, {\it i.e.}\ 6 scalar fields, 4 spin 1/2
fields and 1 vector field; and furthermore, that these fields obey
the usual classical equations of motion.

The same multiplet can also be described in $M=3$ superspace by an irreducible
$M=3$ superspace field strength superfield obeying similar
constraints to the $N=4$ ones but with \eq{dotW} replaced by
 \be
 \bar \nabla_{\adt}^i W_{jk}=-\delta_{[j}{}^i
 \bar \nabla_{\adt}^l W_{k]l}\ .\label{n3constr}
 \ee
Note that the $M=3$ $W_{ij}$ transforms under the complex
3-dimensional representation of the $SU(3)$ internal symmetry
group and is no longer subject to a self-duality condition. This
multiplet has exactly the same field content as the $N=4$
multiplet and is again on-shell. However, this version can be
extended off-shell using harmonic superspace techniques whereas it
is not known how to do this while maintaining manifest $N=4$
supersymmetry.

We give an outline of the off-shell $M=3$ harmonic superspace formalism for the $N=4$
theory in the Appendix. The details of this off-shell theory are not essential for the
superspace power-counting argument which follows, however. This is because it is
sufficient to examine the leading counterterm from the old-style analysis, so
we can simply look at the possible counterterms that can be constructed
using the $M=3$ superfield $W_{ij}$. On the other hand, the
existence of the off-shell harmonic superspace theory (for which a full quantum
formalism can indeed be elaborated\footnote{An example of a full quantum formalism using
harmonic superspace, including the derivation of non-renormalisation theorems, was given
in the context of $D=2$ $(4,0)$ nonlinear supersymmetric sigma models in 
\cite{sokstel}. Quantisation using the harmonic $M=3$ formalism of $D=4$, $N=4$ SYM was
carried out in \cite{delducMcC}.}), assures us that counterterms need also to
be expressible as full superspace integrals in $M=3$ superspace.

The old power-counting rules, assuming that only one-half of the supersymmetry is
preserved, predict that the simplest Lagrangian counterterm would be of the form $\int\,
d^8\theta W^4\sim F^4$ (where $F$ is the spacetime Yang-Mills field strength tensor).
However, if we now make use of the harmonic superspace formalism, we expect the lowest
order counterterms to be of the form $\int\,d^{12}\theta W^4 \sim \del^2 F^4$ where
$\del$ denotes a spacetime derivative. This is under the assumption that we are
quantising using superfields covariant with respect to $D=4$ Lorentz symmetry, even
though the actual spacetime dimension is higher.

Let us now consider the maximal SYM theory in $D=6$. In the old analysis, this was to be
quantised using ordinary $D=6$, $M=1$ superfields ($\leftrightarrow$ $D=4$, $M=2$, {\it
i.e.}\ 8 supercharges). The $D=6$, $M=1$ SYM field strength is a spinor $W^{\alpha
i},i=1,2$, and the off-shell counterterms must, as we have reviewed above, be expressible as
gauge-invariant
$M=1$ superspace integrals. Thus, the old $D=6$ predictions for maximal SYM theory is 
$\int\,d^8\theta\, (W^{\alpha i})^4 \sim \del^2 F^4$. Hence, in the $D=6$ case there is
no change between the old prediction based on preserving 8
supersymmetries with $D=6$ Lorentz covariance and the new prediction based on
preserving 12 supersymmetries with $D=4$ Lorentz covariance; either way, one obtains a
prediction of 3 loops for  the first $D=6$ divergence. What has happened in $D=6$ is that
the requirements of $D=6$ Lorentz plus gauge invariance and of 8-supercharge manifest
supersymmetry coincide with those of gauge and $D=4$ Lorentz invariance and 12-supercharge
supersymmetry.

Now consider the case of $D=5$ maximal SYM. In the old analysis, this was to be quantised
using $D=5$, $M=1$ superfields (again $\leftrightarrow$ $D=4$, $M=2$, {\it i.e.}\ 8
supercharges). The $D=5$, $M=1$ SYM field strength superfield is a scalar $W$, however.
Thus, one would expect to find a divergence of the form $\int\,d^8\theta\, W^4 \sim F^4$.
The new 12-supercharge harmonic superspace prediction improves this to $\del^2 F^4$,
exactly  in agreement with the cutting-rule results of \cite{bernetal}. Note that the
harmonic superspace prediction also reproduces exactly the bound one obtains by assuming
that $D=6$ gauge invariance is still somehow active in $D=5$ as considered in
\cite{karpacz}. Although there is no particular reason to believe that restrictions from
higher-dimensional gauge invariance should continue to be applicable after dimensional
reduction, the new 12-supercharge harmonic superspace analysis is robust, and should apply
to the analysis of all quantum corrections in the theory.

To summarise, let us compare in various spacetime dimensions the predictions of the
old ordinary superspace Feynman rules to the new ones from harmonic superspace, the latter
agreeing fully with the cutting-rule results of \cite{bernetal}. Table \ref{tab1}
gives the results from the old analysis. 

\begin{table}[ht]
\centering
\begin{tabular}{|l|c|c|c|c|c|c|}
\phantomthreeptline\nhline
Dimension&10&8&7&6&5&4\\
loop $L$&1&1&2&3&4&$\infty$\\
gen.\ form&$\partial^2F^4$&$F^4$&$\partial^2F^4$&$\partial^2F^4$&$F^4$&finite\\
\nhline
\nhline\phantomthreeptline
\end{tabular}
\caption{Maximal SYM divergence expectations from 8-supercharge ordinary superspace
Feynman rules\label{tab1}}
\end{table}

These predictions can be compared with those shown in Table \ref{tab2} derived from
the new harmonic superspace analysis, agreeing fully with the unitarity cutting-rule
results of \cite{bernetal}.

\begin{table}[ht]
\centering
\begin{tabular}{|l|c|c|c|c|c|c|}
\phantomthreeptline\nhline
Dimension&10&8&7&6&5&4\\
loop $L$&1&1&2&3&6&$\infty$\\
gen.\
form&$\partial^2F^4$&$F^4$&$\partial^2F^4$&$\partial^2F^4$&
$\partial^2F^4$&finite\\
\nhline
\nhline\phantomthreeptline
\end{tabular}
\caption{Maximal SYM divergence expectations from 12-supercharge harmonic superspace
Feynman rules or from cutting rules.\label{tab2}}
\end{table}

\section{Supergravity}

In order to study supergravity counterterms for quantisation about flat space, it is
sufficient to work at the linearised level.  Linearised $N$-extended
supergravity with $N\geq 5$ is described by a field strength
superfield, $W_{ijkl}$, which is totally antisymmetric in the
$SU(N)$ indices $i,j=1,\ldots N$, and obeys the constraints
 \bea
 D_{\alpha i} W_{jklm}&=& D_{\alpha [i} W_{jklm]}\\
 \bar D_{\adt}^i W_{jklm}&=&-{4\over (N-3)}\delta^i_{[j}
 \bar D_{\adt}^n W_{klm]n}\ .
 \eea
In addition, in the $N=8$ theory, $W$ satisfies a self-duality condition
 \be
 \bar W^{ijkl}={1\over4!}\epsilon^{ijklmnpq} W_{mnpq}\ .
 \ee

The component fields contained in these superfields are precisely
those of the supergravity multiplets and the above constraints imply that they all
satisfy the corresponding linearised supergravity field equations. As in the SYM case,
the maximal theory can be described by either an irreducible $N=8$ or $N=7$ multiplet.

Under the old rules it was assumed that the maximal theory could
be quantised preserving $D=4$, $M=4$ supersymmetry; in this case, the
lowest order counterterm that one can construct is at the 3 loop order and has
the form $\int\,d^{16}\theta\,W^4 \sim R^4$, where $R$ is the
spacetime curvature. This was shown to be fully $N=8$
supersymmetric in \cite{kallosh} and was shown to have full $SU(8)$ internal symmetry
in \cite{superactions} \footnote{This action counterterm can be written in a very 
concise form in a certain harmonic superspace \cite{hh}; in this version all the symmetries are manifest.}. Now let us suppose, in analogy to the Yang-Mills case, that the
theory can be quantised in a harmonic superspace formalism preserving $M=7$ supersymmetry.
In this case the lowest counterterm that would be allowed would be $\int\,
d^{28}\theta\, W^4 \sim \del^6 R^4$. This corresponds to a 6 loop counterterm. In
fact, if $M\geq 4$ four-dimensional supersymmetries can be
preserved, the expected lowest order counterterm would be
$\int\,d^{4M}\theta\, W^4\sim \del^{2M-8} R^4$.

The results of \cite{bernetal} indicate that the onset of divergences in $N=8$
supergravity occurs at the earliest at 5 loops, and this suggests
that there must be an off-shell formulation of the theory with at
least $M=6$ supersymmetry (the 5 loop counterterm is of the form $\del^4
R^4$). However, as we have mentioned above, the analogy with the
maximal Yang-Mills theory suggests that the $N=8$ theory may indeed admit
an off-shell formulation with $M=7$ supersymmetry. If this is the
case, it will be interesting to see if the methods of \cite{bernetal} can be
extended to confirming this explicitly. Alternatively, it would be
interesting to see if harmonic superspace methods can be developed to
find the off-shell theory. Results of the unitarity cutting-rule analysis might serve as a
mathematical ``experiment'' revealing the possibility of an unknown off-shell superspace
formalism for maximal supergravity. This could also have an important impact on the study
of quantum M-theory.

\section*{Appendix: N=3 SYM in harmonic superspace}

The constraints of $N=3$ SYM in ordinary superspace are
 \bea
 F_{\alpha i \beta j}&=&\epsilon_{\alpha\beta} W_{ij} \label{undund}\\
 F_{\alpha i \bdt}^{\phantom{\alpha i}j}&=&0 \label{unddot}
 \eea
where $F$ is the superspace Yang-Mills field strength tensor and
the components which are primarily constrained are those with bi-spinorial
indices. The Bianchi identities then imply constraints on other components, and in
particular imply that the field $W_{ij}$
satisfies the constraints \eq{n3constr} and also that the field equations
of the component fields must be satisfied.

These constraints can be interpreted as integrability conditions
in a harmonic superspace context. To do this, one enlarges the
superspace by adjoining to it an internal manifold of the form
\footnote{This is the same
space as $U(1)\xz U(1)\bsh SU(3)$ which was used in \cite{gikosN4}} 
$U(1)\xz U(1)\xz U(1)\bsh U(3):=H\bsh U(3)$. 
This space is a compact
complex manifold with complex dimension 3. We shall follow the
standard practice of working with fields defined on the group
$U(3)$; this is equivalent to working on the coset if all the
fields are taken to be equivariant with respect to the isotropy
group $H$. We denote an element of $U(3)$ by $u_I{}^i$ and its
inverse by $(u^{-1})_i{}^I$ where the capital indices are acted on
by the isotropy group and the small indices by $U(3)$. Using $u$
and its inverse we can convert $U(3)$ indices to $H$ indices and
vice versa. Thus we have
 \bea
 D_{\alpha I}&:=&u_I{}^i D_{\alpha i}\nn\\
 \bar D_{\adt}^I &:=&\bar D_{\adt}^i (u^{-1})_i{}^I\ .
 \eea
In order to differentiate with respect to the internal manifold we
introduce the right-invariant vector fields $D_I{}^J$. They
satisfy the constraints $D_I{}^I=0$ and $\bar D^I{}_J=-D_J{}^I$.
They also satisfy the commutation relations of $\gu(3)$. The
diagonal derivatives correspond to the isotropy algebra while the
remainder fall into two complex conjugate sets:
$(D_1{}^2,D_1{}^3,D_2{}^3)$ and $(D_2{}^1,D_3{}^1,D_3{}^2)$. The
former three derivatives can be thought of as the components of
the $\bar\del$ operator on the coset and the latter as the
components of the conjugate operator $\del$.

A superfield that is annihilated by $(D_1{}^2,D_1{}^3,D_2{}^3)$
is called H-analytic (H for harmonic), and will have a short harmonic expansion on
the coset since this space is a compact complex manifold. A
superfield which is annihilated by $(D_{\alpha 1},\bar D_{\adt}^3)$ is
called G-analytic (G for Grassmann), and a superfield which
annihilated by both sets of operators is called CR-analytic, or
just analytic for short.

The SYM constraints can now be interpreted as integrability
conditions in harmonic superspace. They correspond to the
vanishing of the SYM curvature in the spinorial directions $\left(\alpha
1,\stackrel{3}{\adt}\right)$. These constraints are
 \be
 F_{\alpha 1\beta 1}=F_{\adt\bdt}^{33}=F_{\alpha 1\bdt}^{\phantom{\alpha i}3}=0
\label{intconds}
 \ee

At this stage we have not yet extended the gauge theory into the new directions so that 
the $F_{\alpha i\beta j}$ {\it etc.}\ do not depend on the
harmonic coordinates. It therefore follows that the constraints (\ref{intconds}) imply (\ref{undund},\ref{unddot})
above, and these in turn imply the equations of
motion. We can take the theory off-shell by introducing a
connection for the harmonic directions and by allowing the components
of the curvature in these directions to be non-vanishing. In
this case, the constraints \eq{intconds} no longer imply the original
constraints (\ref{undund},\ref{unddot}). To get the equations of motion we therefore
only need to find an action which will imply that the curvature
should vanish in the harmonic directions. Since this space has
three complex dimensions one can use a Chern-Simons action for
this. Remarkably, all of the dimensions and internal charges work
out just right for this to work.

The resulting action is
\be
I=\int\,d\mu^{11}_{33}\, Q^{(3)33}_{\phantom{(3)}11}
\ee
where the superscripts indicate the $(U(1))^3$ charges and the measure $d\mu$ is defined by

\be
d\mu^{11}_{33}:= d^4x\, du\, \left(D_2\,D_3\,\bar D^1\,\bar D^2\right)^2
\ee

Here $du$ is the usual measure for the coset space while $(D_2)^2:=D_{\alpha 2} D^{\alpha 2}$ {\it etc}.
$Q^{(3)}$ is the Chern-Simons 3-form defined in the usual way (on the whole of harmonic superspace)
by $dQ^{(3)}={\rm tr} (F\wedge F)$. 
The component of $Q^{(3)}$ which appears in the action is the component in the anti-holomorphic
harmonic direction, 
$Q_{\phantom{(3)}1\phantom{2}, 1\phantom{3}, 2}^{{(3)}\phantom{1}2\phantom{,1}3\phantom{,2}3}
:=Q^{(3)33}_{\phantom{(3)}11}$. 
This expression
can be written explicitly in terms of the connection in the harmonic directions which has
components $(A_1{}^2,A_1{}^3,A_2{}^3)$. The above constraints, together with the fact that 
the mixed harmonic/superspace curvatures are also zero imply that these components of 
the connection are G-analytic so that the action is manifestly supersymmetric.

\section*{Acknowledgements}

We would like to thank Lance Dixon for discussions, and the Isaac Newton Institute for
hospitality during the early stages of this work.

This work was supported in part by PPARC under SPG grant PPA/G/S/1998/00613.

\end{document}